\documentclass[12pt]{article}
\usepackage{graphicx}
\def\beq{\begin{equation}}
\def\eeq{\end{equation}}
\def\bea{\begin{eqnarray}}
\def\eea{\end{eqnarray}}

\def\sss{\scriptscriptstyle}
\def\bd{B_d^0}
\def\bdbar{{\bar B}^0_d}
\def\bs{B_s^0}
\def\bsbar{{\bar B}^0_s}
\def\barp{{\raise.35ex\hbox
{${\sss (}$}}---{\raise.35ex\hbox{${\sss )}$}}}
\def\bdbarp{\hbox{$B_d$\kern-1.4em\raise1.4ex\hbox{\barp}}}
\def\bsbarp{\hbox{$B_s$\kern-1.4em\raise1.4ex\hbox{\barp}}}
\def\ks{K_{\sss S}}

\def\roughly#1{\mathrel{\raise.3ex\hbox
{$#1$\kern-.75em\lower1ex\hbox{$\sim$}}}}
\def\lsim{\roughly<}

\def\bra#1{\langle#1|}
\def\ket#1{|#1\rangle}
\def\barpk{{\raise.35ex\hbox
{${\sss (}$}}--{\raise.35ex\hbox{${\sss )}$}}}
\def\kbarp{\hbox{$K$\kern-0.9em\raise1.4ex\hbox{\barpk}}}
\def\epjc#1#2#3{{\it Eur.\ Phys.\ J.}\ {\bf C#1} (#2) #3}

\def\npb#1#2#3{{\it Nucl.\ Phys.} {\bf B#1} (#2) #3}
\def\plb#1#2#3{{\it Phys.\ Lett.} {\bf #1B} (#2) #3}

\def\prd#1#2#3{{\it Phys.\ Rev.} {\bf D#1} (#2) #3}
\def\newprd#1#2#3{{\it Phys.\ Rev.} {\bf D#1}: #3 (#2)}

\def\prep#1#2#3{{\it Phys.\ Rep.} {\bf #1} (#2) #3}
\def\prl#1#2#3{{\it Phys.\ Rev.\ Lett.} {\bf #1} (#2) #3}

\def\zpc#1#2#3{{\it Zeit.\ Phys.} {\bf C#1} (#2) #3}

\newread\epsffilein 
\newif\ifepsffileok 
\newif\ifepsfbbfound 
\newif\ifepsfverbose 
\newdimen\epsfxsize 
\newdimen\epsfysize 
\newdimen\epsftsize 
\newdimen\epsfrsize 
\newdimen\epsftmp 
\newdimen\pspoints 
\def\epsfbox#1{\global\def\epsfllx{72}\global\def\epsflly{72}%
 \global\def\epsfurx{540}\global\def\epsfury{720}%
 \def\lbracket{[}\def\testit{#1}\ifx\testit\lbracket
 \let\next=\epsfgetlitbb\else\let\next=\epsfnormal\fi\next{#1}}%
\def\epsfgetlitbb#1#2 #3 #4 #5]#6{\epsfgrab #2 #3 #4 #5 .\\%
 \epsfsetgraph{#6}}%
\def\epsfnormal#1{\epsfgetbb{#1}\epsfsetgraph{#1}}%
\def\epsfgetbb#1{%
\openin\epsffilein=#1
\ifeof\epsffilein\errmessage{I couldn't open #1, will ignore it}\else
 {\epsffileoktrue \chardef\other=12
 \def\do##1{\catcode`##1=\other}\dospecials \catcode`\ =10
 \loop
 \read\epsffilein to \epsffileline
 \ifeof\epsffilein\epsffileokfalse\else
 \expandafter\epsfaux\epsffileline:. \\%
 \fi
 \ifepsffileok\repeat
 \ifepsfbbfound\else
 \ifepsfverbose\message{No bounding box comment in #1; using defaults}\fi\fi
 }\closein\epsffilein\fi}%
\def\epsfclipstring{}
\def\epsfsetgraph#1{%
 \epsfrsize=\epsfury\pspoints
 \advance\epsfrsize by-\epsflly\pspoints
 \epsftsize=\epsfurx\pspoints
 \advance\epsftsize by-\epsfllx\pspoints
 \epsfxsize\epsfsize\epsftsize\epsfrsize
 \ifnum\epsfxsize=0 \ifnum\epsfysize=0
 \epsfxsize=\epsftsize \epsfysize=\epsfrsize
 \epsfrsize=0pt
 \else\epsftmp=\epsftsize \divide\epsftmp\epsfrsize
 \epsfxsize=\epsfysize \multiply\epsfxsize\epsftmp
 \multiply\epsftmp\epsfrsize \advance\epsftsize-\epsftmp
 \epsftmp=\epsfysize
 \loop \advance\epsftsize\epsftsize \divide\epsftmp 2
 \ifnum\epsftmp>0
 \ifnum\epsftsize<\epsfrsize\else
 \advance\epsftsize-\epsfrsize \advance\epsfxsize\epsftmp \fi
 \repeat
 \epsfrsize=0pt
 \fi
 \else \ifnum\epsfysize=0
 \epsftmp=\epsfrsize \divide\epsftmp\epsftsize
 \epsfysize=\epsfxsize \multiply\epsfysize\epsftmp
 \multiply\epsftmp\epsftsize \advance\epsfrsize-\epsftmp
 \epsftmp=\epsfxsize
 \loop \advance\epsfrsize\epsfrsize \divide\epsftmp 2
 \ifnum\epsftmp>0
 \ifnum\epsfrsize<\epsftsize\else
 \advance\epsfrsize-\epsftsize \advance\epsfysize\epsftmp \fi
 \repeat
 \epsfrsize=0pt
 \else
 \epsfrsize=\epsfysize
 \fi
 \fi
 \ifepsfverbose\message{#1: width=\the\epsfxsize, height=\the\epsfysize}\fi
 \epsftmp=10\epsfxsize \divide\epsftmp\pspoints
 \vbox to\epsfysize{\vfil\hbox to\epsfxsize{%
 \ifnum\epsfrsize=0\relax
 \includegraphics{#1}%
 \else
 \epsfrsize=10\epsfysize \divide\epsfrsize\pspoints
 \includegraphics{#1}%
 \fi
 \hfil}}%
\global\epsfxsize=0pt\global\epsfysize=0pt}%
 {\catcode`\%=12 \global\let\epsfpercent=
\long\def\epsfaux#1#2:#3\\{\ifx#1\epsfpercent
 \def\testit{#2}\ifx\testit\epsfbblit
 \epsfgrab #3 . . . \\%
 \epsffileokfalse
 \global\epsfbbfoundtrue
 \fi\else\ifx#1\par\else\epsffileokfalse\fi\fi}%
\def\epsfempty{}%
\def\epsfgrab #1 #2 #3 #4 #5\\{%
\global\def\epsfllx{#1}\ifx\epsfllx\epsfempty
 \epsfgrab #2 #3 #4 #5 .\\\else
 \global\def\epsflly{#2}%
 \global\def\epsfurx{#3}\global\def\epsfury{#4}\fi}%
\def\epsfsize#1#2{\epsfxsize}
\begin{document}

\begin{flushright}  
UdeM-GPP-TH-02-104\\
\end{flushright}

\begin{center} 

{\large \bf \centerline{\boldmath $b$ Physics Beyond the Standard
Model\footnote{Talk given at {\it Beauty 2002}, Santiago de
Compostela, Spain, June 2002.}}}
\vspace*{0.5cm}
{\large David
  London\footnote{email: london@lps.umontreal.ca}} \vskip0.3cm
{\it  Laboratoire Ren\'e J.-A. L\'evesque, Universit\'e de
  Montr\'eal,} \\
{\it C.P. 6128, succ.\ centre-ville, Montr\'eal, QC, Canada H3C 3J7} \\
\vskip0.3cm
\bigskip
(\today)
\vskip0.3cm
{\Large Abstract\\}
\vskip3truemm
\parbox[t]{\textwidth} {I review the signals for New Physics in
CP-violating measurements in $B$ and $\Lambda_b$ decays. I also
discuss ways of identifying this New Physics, should such a signal be
found.}
\vskip0.3cm
\end{center}
\baselineskip=14pt

For the past decade or so, there has been an enormous amount of
effort, both theoretical and experimental, devoted to the study of $b$
physics. The main goal, as always, is to find physics beyond the
standard model (SM). In this review, I will address two questions:
\begin{enumerate}

{\item What are signals of New Physics?}

{\item If such a signal is found, can we identify the New Physics?}

\end{enumerate}
I will focus principally, but not exclusively, on measurements which
can be made at hadron colliders. Also, I will concentrate on
measurements of CP violation.

In order to detect New Physics, we must observe a deviation from the
SM prediction for some process. However, {\it all} such predictions
have some theoretical input, and the uncertainty on this input limits
our ability to deduce the presence of New Physics. In my discussion
below, I will rate the various signals of New Physics using the
ever-popular ``star system'' \cite{Hoecker} to indicate the size of
the theoretical uncertainty:
\begin{itemize}

\item $\star\star\star$ $\Longrightarrow$ theoretical uncertainty
$\lsim 1\%$,

\item $\star\star$ $\Longrightarrow$ theoretical uncertainty $\lsim
5\%$,

\item $\star$ $\Longrightarrow$ theoretical uncertainty $\lsim 25\%$.

\end{itemize}
(This is inspired by Cabibbo angle counting: $\lambda \sim 25\%$,
$\lambda^2 \sim 5\%$, $\lambda^3 \sim 1\%$.)

I begin the discussion by reviewing the predictions of the SM, along
with the size of the theoretical uncertainty. Note that the following
list of predictions is long, but not comprehensive. Any deviation from
the SM prediction indicates the presence of New Physics.

\begin{itemize}

\item ${\cal A}_{CP}^{dir} (B \to \Psi K) = 0$. The decay $B \to \Psi
K$ (charged or neutral) has effectively only one weak decay amplitude.
In addition to the tree amplitude, there may be a penguin
contribution, but (i) it is expected to be small, and (ii) its weak
phase is essentially the same as that of the tree. The direct CP
asymmetry is therefore predicted to vanish in the SM. \quad
$\star\star\star$

\item ${\cal A}_{CP}^{mix} (\bd(t)\to \Psi \ks) = {\cal A}_{CP}^{mix}
(\bd(t)\to \phi \ks)$. The decay $\bd\to \phi \ks$ is pure $b\to s$
penguin, which is dominated by an internal $t$-quark (CKM matrix
elements $V_{tb}^* V_{ts}$). In the Wolfenstein parametrization of the
CKM matrix \cite{Wolfenstein}, $V_{tb}^* V_{ts}$ has only a small
[$O(\lambda^2)$] imaginary piece, so that the penguin decay amplitude
is approximately real, as is the amplitude for $\bd\to \Psi \ks$.
Both CP asymmetries therefore measure $\sin 2\beta$ in the SM. Any
deviation from this result indicates the presence of New Physics in
the $b\to s$ penguin amplitude \cite{NPpenguins}. \quad $\star\star$

\item ${\cal A}_{CP}(\bd(t)\to D^* \rho)$ measures $2\beta + \gamma$.
The angular analysis of this decay mode allows one to extract the
quantity $2\beta + \gamma$ \cite{D*rho}. This may well be the second
function of CP phases, after $\sin 2\beta$, to be measured at
$B$-factories. \quad $\star\star\star$

\item ${\cal A}_{CP}^{mix}(\bs(t) \to \Psi \phi) \simeq 0$. In the SM,
this CP asymmetry probes ${\rm arg}(V_{cb}^* V_{cs} V_{tb} V_{ts}^*)$.
As mentioned above, $V_{ts}$ has a small [$O(\lambda^2)$] imaginary
piece in the Wolfenstein parametrization, so a nonzero asymmetry is
expected at the several percent level. If a larger asymmetry is
measured, this probably indicates New Physics in $\bs$--$\bsbar$
mixing. \quad $\star\star$

\item ${\cal A}_{CP}(B^\pm \to D K^\pm) = {\cal A}_{CP}^{mix}(\bs(t)
\to D_s^\pm K^\mp)$. In the SM, both of these CP asymmetries probe the
weak phase $\gamma$ \cite{BDK,BsDsK}. Any discrepancy in the values of
$\gamma$ obtained in these methods points to the presence of New
Physics in $\bs$--$\bsbar$ mixing. \quad $\star\star$

\item Inclusive ${\cal A}_{CP}^{dir} (b \to s\gamma) = 0$. In order to
have a direct CP asymmetry, one requires two decay amplitudes with
different weak and strong phases. For the decay $b \to s\gamma$,
rescattering effects must be significant for this to occur. However,
in the SM, such effects are tiny \cite{Soares}, so that the direct
asymmetry effectively vanishes. \quad $\star\star\star$

\item Exclusive ${\cal A}_{CP}^{mix} (b \to s\gamma) = 0$. The photon
emitted in the decay $b \to s\gamma$ is predominantly left-handed,
while that emitted in the CP-conjugate decay is right-handed. Thus,
one does not have the same final state for $\bd$ and $\bdbar$ decays,
so that the indirect CP asymmetry for exclusive states is expected to
vanish \cite{AGS}. Of course, this is true only to the extent that the
photon helicity is purely left-handed or right-handed. In fact, there
are corrections of $O(2 m_s/m_b) \simeq 5\%$, so that this is the size
of the asymmetry expected in the SM. Anything larger indicates New
Physics. \quad $\star\star$

\item There are many techniques for getting CKM phase information
using flavour $SU(3)$ symmetry \cite{SU3}. Generally, these are
1 $\star$ methods, since $SU(3)$-breaking effects are typically of
order $f_K/f_\pi \sim 25\%$. In order to detect New Physics using such
methods, the effects must therefore be larger than this theoretical
uncertainty.

\item One exception is a method involving $B^0_{d,s} \to K^{(*)} {\bar
K}^{(*)}$ decays, which can be used to obtain the CP phase $\alpha$
\cite{BKKbar}. In this case, one extracts $\alpha$ using a double
ratio in which the $SU(3)$-breaking effects largely cancel, thus
reducing the theoretical uncertainty considerably. \quad $\star\star$

\item Suppose that ${\cal A}_{CP}^{mix}(\bs(t) \to \Psi \phi)$ is
actually measured, and the CKM phase $\chi$ ($\sim$ 2 -- 5\%)
extracted. Within the 3-generation SM, one expects \cite{3genSM}
\beq
\sin\chi = \left\vert \frac{V_{us}}{V_{ud}} \right\vert^2
\frac{\sin\beta \sin(\gamma-\chi)}{\sin(\beta+\gamma)} ~.
\eeq
Any violation of this relation implies the presence of New Physics.
\quad $\star\star\star$

\item One can look for an inconsistency between the unitarity triangle
as constructed from measurements of the angles and that constructed
from measurements of the sides. However, this method is limited by the
large theoretical uncertainties in the extraction of the sides. \quad
$\star$

\item One can look for an inconsistency between measurements of the
angles and exclusive hadronic rates. For example, indirect constraints
on $\gamma$ imply that $\gamma < 90^\circ$, while early studies of $B
\to \pi K$ decays seemed to indicate that $\gamma > 90^\circ$
\cite{Hou}. Of course, the hadronic uncertainties are very large in
such methods, so it is difficult to conclude that New Physics is
present. \quad $\star$

\end{itemize}
As stated earlier, the above is only a partial list of possible
signals of New Physics. Because of space limitations, I cannot give
more than a cursory description of each. However, below I discuss in
more detail two additional observables which are perhaps less
well-known.

Suppose a New-Physics amplitude contributes to (charged or neutral) $B
\to \Psi K$ decays. How would we detect it? As mentioned above, the
obvious answer is to measure the direct CP asymmetry. However,
\beq
{\cal A}_{CP}^{dir} (B \to \Psi K) \sim A_{\sss SM} A_{\sss NP}
\sin\phi \sin\Delta ~,
\eeq
where $\phi$ and $\Delta$ are the relative weak and strong phases,
respectively. Unfortunately, from this expression one sees that if
$\Delta = 0$, the direct CP asymmetry will vanish, and the New Physics
will remain hidden.

The situation can be improved by considering instead the decay $B\to
\Psi K^*$ \cite{PsiK*}. The time-dependent decay rate can be written
in the helicity basis as
\beq
\Gamma(\bd(t)\to \Psi K^*) \sim \sum_{\lambda\sigma} \left[
\Lambda_{\lambda\sigma} + \Sigma_{\lambda\sigma} \cos\Delta m t -
\rho_{\lambda\sigma} \sin\Delta m t \right] g_\lambda g_\sigma ~,
\eeq
where $\lambda,\sigma = 0,\|,\perp$. The important observable here is
$\Lambda_{\perp i}$, which can be written schematically as
\beq
\Lambda_{\perp i} \sim (A_{\sss SM}^\perp A_{\sss NP}^i - A_{\sss
SM}^i A_{\sss NP}^\perp) \sin\phi \cos(\Delta_{\perp i}) ~,~i = 0,\|~.
\eeq
Note the appearance of the $\cos(\Delta_{\perp i})$ factor. It is this
term which appears (and not a $\sin(\Delta_{\perp i})$ term) due to
the interference of CP-even and CP-odd helicities. The key point is
that even if the strong phases are zero, $\Delta_{\perp i} = 0$,
$\Lambda_{\perp i}$ will not vanish. Thus, this observable is
complementary to ${\cal A}_{CP}^{dir} (B \to \Psi K)$, and it will be
important to measure both of these quantities to test for the presence
of New Physics. Note also that the theoretical uncertainty is tiny, so
that $\Lambda_{\perp i}$ is a 3 $\star$ observable.

Furthermore, this holds equally for $\bd(t)\to \Psi K^*$ and for
$B^\pm \to \Psi K^\pm$. That is, one can simply add all charged and
neutral $B$ decays together -- no tagging or time-dependent
measurements are needed to obtain $\Lambda_{\perp i}$, making it a
particularly interesting quantity from an experimental point of view.

All of the methods mentioned above deal with $B$ mesons. However,
there are also useful tests for New Physics using $B$ baryons. In
particular, one can look at T-violating triple-product correlations in
charmless $\Lambda_b$ decays \cite{BDL1}.

Triple-product (TP) correlations take the form $\vec v_1 \cdot (\vec
v_2 \times \vec v_3)$, where each $v_i$ is a spin or momentum. TP
correlations are odd under T, which implies, by the CPT theorem, that
they are also odd under CP. By measuring a nonzero value of
\beq
A_T \equiv \frac{\Gamma (\vec v_1 \cdot (\vec v_2 \times \vec v_3)>0) - 
\Gamma (\vec v_1 \cdot (\vec v_2 \times \vec v_3)<0)}{
\Gamma (\vec v_1 \cdot (\vec v_2 \times \vec v_3)>0) + 
\Gamma (\vec v_1 \cdot (\vec v_2 \times \vec v_3)<0)} ~,
\eeq
one obtains a signal for a nonzero TP correlation. However, there is a
complication: strong phases can produce a nonzero value of $A_T$, even
if there is no CP violation (i.e.\ if the weak phases are zero). In
order to be sure that one is truly probing T and CP violation, the
value of $A_{\sss T}$ must be compared with that of ${\bar A}_{\sss
T}$, which is the T-odd asymmetry measured in the CP-conjugate decay
process.

Consider the decays $\Lambda_b \to F_1 P$ and $F_1 V$, where $F_1$ is
a fermion ($p$, $\Lambda$, ...), $P$ is a pseudoscalar ($K^-$, $\eta$,
...), and $V$ is a vector ($K^{*-}$, $\phi$, ...). For $\Lambda_b \to
F_1 P$, there is only one possible triple product: $\vec p_{F_1} \cdot
(\vec s_{F_1} \times \vec s_{\Lambda_b})$. On the other hand, in
$\Lambda_b \to F_1 V$ decays, one has 3 spins and 1 independent
momentum. This implies that there are 4 possible TP's. In all cases,
we would like to know the expectations for the sizes of these TP's in
the SM.

One can analyze $\Lambda_b$ decays using factorization \cite{BDL1}:
\beq
A(\Lambda_b \to F_1 P/V) = \sum_{{\cal O},{\cal O}'} \langle P/V \vert
{\cal O} \vert 0 \rangle \langle F_1 \vert {\cal O}' \vert \Lambda_b
\rangle ~.
\eeq
For $\Lambda_b \to F_1 P$ one can then write
\beq
{\cal M}_P = if_P p_P^{\mu} \, \bra{F_1} \bar{u} \gamma_\mu(1-\gamma_5)
b\ket{\Lambda_b} X_P + i f_P p_P^{\mu} \, \bra{F_1}\bar{u}
\gamma_\mu(1+\gamma_5) b\ket{\Lambda_b} Y_P ~,
\eeq
where $X_P$ and $Y_P$ are functions of CKM matrix elements, Wilson
coefficients and masses. Similarly, for $\Lambda_b \to F_1 V$, one has
\beq
{\cal M}_V = m_V g_V \varepsilon_V^{*\mu} \bra{F_1} \bar{u}
\gamma_\mu(1-\gamma_5) b\ket{\Lambda_b} X_V + m_V g_V
\varepsilon_V^{*\mu} \bra{F_1}\bar{u} \gamma_\mu(1+\gamma_5)
b\ket{\Lambda_b} Y_V ~.
\label{TPvec}
\eeq
Like any CP-violating signal, in order to have a nonzero
triple-product correlation, one needs two interfering amplitudes.
Thus, from the above expressions, one sees that both $X_P$ and $Y_P$
($X_V$ and $Y_V$) must be nonzero to have TP's in $\Lambda_b \to F_1
P$ ($\Lambda_b \to F_1 V$).

Now, it is easy to see that both $X_P$ and $X_V$ are nonzero in the
SM. After all, all operators in effective hamiltonian involve a
left-handed $b$. The real question is: are $Y_P$ and/or $Y_V$ nonzero
in the SM? The answer is that, for particular $F_1 P$ final states,
one can ``grow'' a right-handed current due to the Fierzing of certain
SM operators. Furthermore, the $Y_P$ in such cases can be sizeable.
However, for $F_1 V$ final states, this doesn't work -- the matrix
elements of the above operators vanish for a final-state
$V$. Therefore $Y_V \simeq 0$ in SM. The end result is that
\cite{BDL1}
\begin{itemize}

\item ${\cal A}_T^{pK} = -18\%$.

\item The triple products for all other decays are expected to be
small, at most $O(1\%)$. These include the final states $p K^{*-}$,
$\Lambda \eta$, $\Lambda\eta'$, $\Lambda\phi$.

\end{itemize}

The bottom line is that many triple-product correlations in
$\Lambda_b$ decays are expected to be tiny in the SM. This suggests
that this is a good place to look for New Physics.

I now turn to the issue of identifying the New Physics. Should a
New-Physics signal be found, precise identification will have to wait
for direct production at high-energy colliders. However, it is still
possible to get a fairly good idea of the type of New Physics just by
studying $B$/$\Lambda_b$ processes.

New Physics can affect $B^0$--${\bar B}^0$ mixing and/or
$B$/$\Lambda_b$ decays. One expects that penguin decays will be most
affected, but there could be New-Physics contributions to tree-level
processes. It is therefore useful to classify the New Physics as
affecting either the $b\to s$ FCNC or the $b\to d$ FCNC.

With this in mind, there are two complementary approaches to
identifying the New Physics:
\begin{enumerate}

\item One can consider various New-Physics models \cite{Bnewphysics}.
For each model, one examines the predicted effects on CP violation and
rare $B$/$\Lambda_b$ decays. By comparing the pattern of effects with
what is observed, one can see if the model is a viable candidate.

\item One can analyze effects using a model-independent (effective
lagrangian) approach. The determination of which New-Physics operators
are or are not present will help in ruling out candidate models of New
Physics.

\end{enumerate}
Below I give examples of each of these approaches.

I begin with a discussion of two specific models of New Physics, and
examine their predictions for various processes in the $B$ system. The
first model involves $Z$-mediated FCNC's: if the $d$, $s$ and
$b$-quarks mix with a vector-singlet down-type quark, flavour-changing
$Z$ couplings are generated \cite{NirSilv}. In particular, one can
have $Zb{\bar d}$ and $Zb{\bar s}$ FCNC couplings, denoted $U_{db}$
and $U_{sb}$. These may be complex, and can contribute to both
$B^0$--${\bar B}^0$ mixing and $B$/$\Lambda_b$ decays.

The constraints on $U_{db}$ and $U_{sb}$ come principally from the
following measurements: $BR(B \to X \mu^+\mu^-) < 5 \times 10^{-5}$
(UA1) \cite{ua1} and $BR(B \to X_s e^+ e^-) \le 1.01 \times 10^{-5}$
(BELLE) \cite{BELLE}. One finds \cite{LerLon}
\beq
\left\vert U_{db} \right\vert \le 0.002 ~~,~~~~
\left\vert U_{sb} \right\vert \le 7.6 \times 10^{-4} ~.
\eeq
With these constraints, $Z$-mediated FCNC's will not significantly
affect the rates for $b\to s$ FCNC's \cite{Hiller}. However, there can
be important effects in $b\to d$ FCNC processes \cite{Bnewphysics}:
$Z$-mediated FCNC's can have
\begin{itemize}

\item significant effects on $\bd$--$\bdbar$ mixing, with or without
new phases,

\item little effect on the rates for gluonic $b\to d$ penguin
processes, such as $\bd\to K^0 {\bar K}^0$,

\item {\it huge} effects on $b\to d \ell^+ \ell^-$,
$\bd\to\ell^+\ell^-$ and $b\to d$ electroweak-penguin decays, such as
$B^+ \to \phi \pi^+$.

\end{itemize}
Thus, should this pattern of New-Physics effects be observed, it would
suggest the presence of $Z$-mediated FCNC's \cite{Branco}.

Another model of New Physics which has been much discussed is
supersymmetry (SUSY). In SUSY models with minimal flavour violation,
all contributions to $B^0$--${\bar B}^0$ mixing and penguin decays are
proportional to the same combination of CKM matrix elements as found
in the SM. As a consequence, there are no new effects in CP-violating
observables, and the extracted values of $\alpha$, $\beta$, $\gamma$
will be the true (SM) values. The unitarity triangle (UT) as
constructed from measurements of the angles will therefore be the SM
UT. On the other hand, there are SUSY contributions to $K^0$--${\bar
K}^0$, $\bd$--$\bdbar$ and $\bs$--$\bsbar$ mixing, so that the
measurements of the sides of the unitarity triangle are not the true
SM values. The UT as constructed from the sides will therefore {\it
not} be the SM UT. Thus, one can detect this type of New Physics by
observing a discrepancy between the two unitarity triangles.

In such minimal SUSY models, the contributions to meson mixing can be
distinguished by a single parameter $f$. Depending on the value of
$f$, the profile of the unitarity triangle will change \cite{AliLon}:
%
\begin{center}
\centerline{\epsfysize 10cm \epsfbox{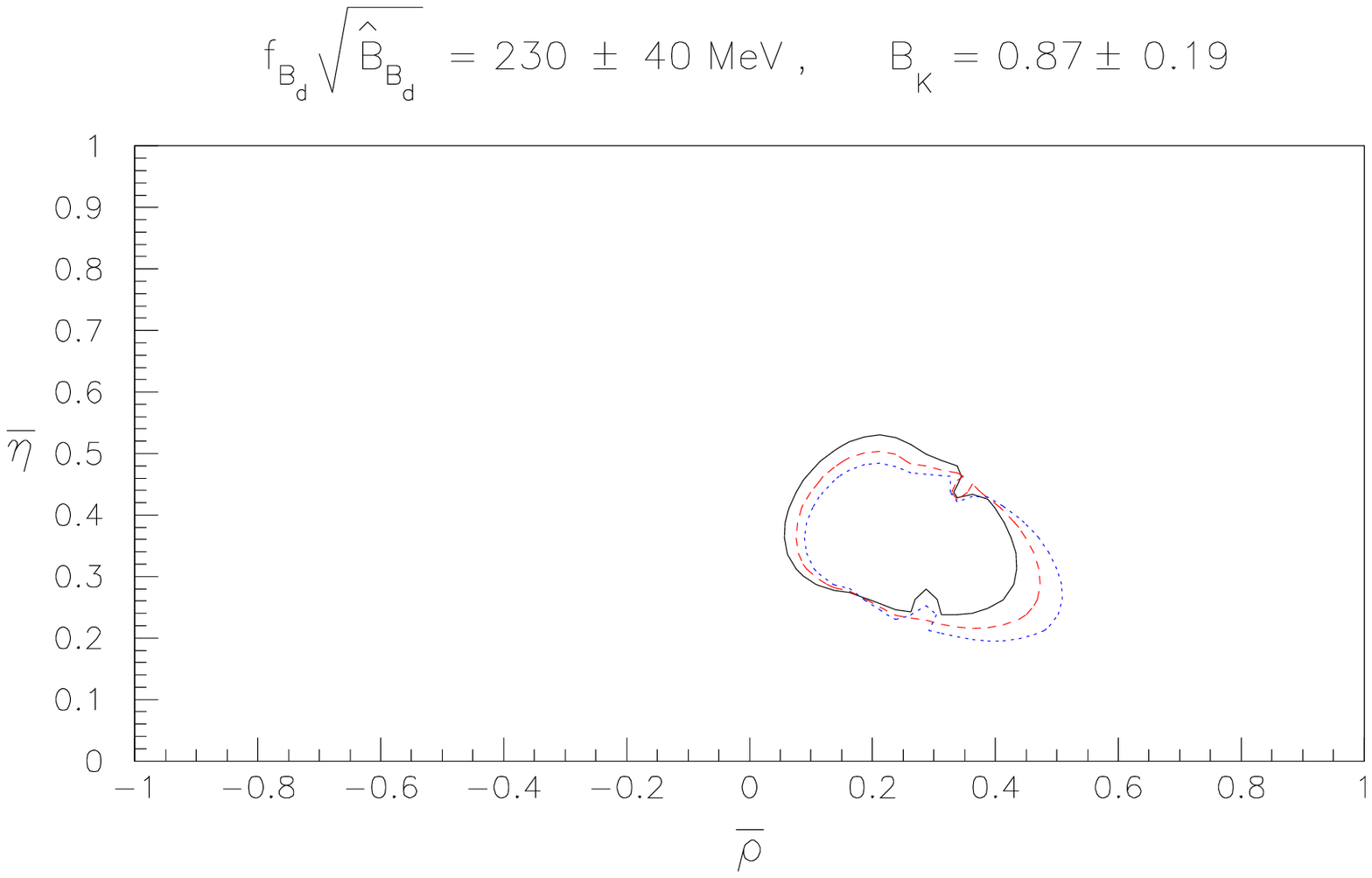}} 
\end{center}
%
(Here the allowed regions correspond to $f=0$ [SM, solid line],
$f=0.25$ [long dashed line], and $f=0.5$ [short dashed line].) The
size of these regions is due principally to the large theoretical
errors of $\sim 20\%$.

In order to detect this type of New Physics, one must distinguish the
UT of the SM ($f=0$) from one with a nonzero value of $f$. From the
above figure, it is obvious that this is will be very difficult to do.
Therefore, unless we can substantially reduce the theoretical
uncertainties, it will be almost impossible to see the effects of
minimal SUSY models in the $B$ system. On the other hand, if one finds
New-Physics effects in CP-violating observables (e.g.\ ${\cal
A}_{CP}^{mix} (\bd(t)\to \Psi \ks) \ne {\cal A}_{CP}^{mix} (\bd(t)\to
\phi \ks)$), then these models can be ruled out.

I now turn to model-independent methods for identifying the New
Physics. Consider first the decay $B\to \Psi K$. One can perform an
isospin decomposition of this decay: since $B$ and $K$ are
isodoublets, and $\Psi$ an isosinglet, the weak hamiltonian has an
$I=0$ and $I=1$ piece, with
\bea
\langle \Psi K^+ \vert {\cal H}_{eff}^{I=0} \vert B^+ \rangle &=&
\langle \Psi K^0 \vert {\cal H}_{eff}^{I=0} \vert \bd \rangle ~, \\
\langle \Psi K^+ \vert {\cal H}_{eff}^{I=1} \vert B^+ \rangle &=&
-\langle \Psi K^0 \vert {\cal H}_{eff}^{I=1} \vert \bd \rangle ~.
\eea

Now, we know that any direct CP asymmetry in this decay is a signal of
New Physics. But what kind of New Physics is it? We can obtain some
information as follows \cite{FM}: define
\beq
{\cal A}_{CP}^+ \equiv {\hbox{${\cal A}_{CP}^{dir}$ in $B^\pm \to \Psi
K^\pm$}} ~~,~~~~
{\cal A}_{CP}^0 \equiv {\hbox{${\cal A}_{CP}^{dir}$ in $\bdbarp \to \Psi
\kbarp$}}
\eeq
and
\beq
S \equiv \frac{1}{2}\left[ {\cal A}_{CP}^0 + {\cal A}_{CP}^+ \right] ~,~~~
D \equiv \frac{1}{2}\left[ {\cal A}_{CP}^0 - {\cal A}_{CP}^+ \right]
~.
\eeq
Note that $S\ne 0$ and $D\ne 0$ are both signals of New Physics.
However, there is a difference between them: $S$ is due to isospin 0
effects, while $D$ is due to isospin 1 effects. Their measurement
gives us model-independent information about the underlying New
Physics.

Finally, I return to triple products in $\Lambda_b$ decays. Above, we
saw that the triple-product asymmetries in $\Lambda_b \to p K^{*-}$
are expected to be very small. Suppose now that we measure a large
triple product. What New Physics could be responsible?

{}From Eq.~(\ref{TPvec}), we see that we need operators which
contribute to the matrix element $\bra{p}\bar{u}
\gamma_\mu(1+\gamma_5) b\ket{\Lambda_b}$. It is straightforward to
write these down. They are \cite{BDL2}:
\beq
{\bar s} (1 + \gamma_5) b \, {\bar u} (1 - \gamma_5) u ~~,~~~~ {\bar
s} \gamma^\mu (1 + \gamma_5) b \, {\bar u} \gamma_\mu (1 + \gamma_5) u
~.
\eeq
The point here is that a significant triple-product signal in
$\Lambda_b \to p K^{*-}$ would (i) tell us that New Physics is
present, and (ii) indicate which New-Physics operators can contribute.
Thus, triple-product asymmetries in $\Lambda_b$ decays can serve as a
diagnostic tool for New Physics.

This same procedure can be applied to other decays such as $\Lambda_b
\to \Lambda \eta$, $\Lambda_b \to \Lambda \phi$, etc. In this way, we
can get a more complete picture of which New-Physics operators are or
are not present \cite{BDL2}.

To sum up: there are many, many signals of New Physics in
$B$/$\Lambda_b$ processes. In addition, there are many ways of
determining which types of New Physics might be responsible for these
signals. It is quite likely that we will have a fairly good idea of
what kind of New Physics is present in these decays.

{\it Let's hope that Nature is kind, and we actually see some evidence
of New Physics!}


\end{document}